\documentclass[twocolumn, dvipsnames]{aastex631}
\usepackage{xspace}

\usepackage{multirow}
\usepackage{hyperref}

\newcommand{\Vsquared}{V$^2$\xspace}

\begin{document}

\title{DESIVAST: Catalogs of Low-Redshift Voids using Data from the DESI Data Release 1 Bright Galaxy Survey}
\author[0009-0001-1962-3924]{Hernan Rincon}\email{hrincon@ur.rochester.edu}
\affiliation{Department of Physics \& Astronomy, University of Rochester, 206 Bausch and Lomb Hall, P.O. Box 270171, Rochester, NY 14627-0171, USA}
\author[0000-0001-5537-4710]{Segev BenZvi}\email{segev.benzvi@rochester.edu}
\affiliation{Department of Physics \& Astronomy, University of Rochester, 206 Bausch and Lomb Hall, P.O. Box 270171, Rochester, NY 14627-0171, USA}
\author[0000-0002-9540-546X]{Kelly Douglass}\email{kellyadouglass@rochester.edu}
\affiliation{Department of Physics \& Astronomy, University of Rochester, 206 Bausch and Lomb Hall, P.O. Box 270171, Rochester, NY 14627-0171, USA}
\author[0000-0001-8101-2836]{Dahlia Veyrat}\email{dveyrat@ur.rochester.edu}
\affiliation{Department of Physics \& Astronomy, University of Rochester, 206 Bausch and Lomb Hall, P.O. Box 270171, Rochester, NY 14627-0171, USA}
\author{Jessica Nicole Aguilar}
\affiliation{Lawrence Berkeley National Laboratory, 1 Cyclotron Road, Berkeley, CA 94720, USA}
\author[0000-0001-6098-7247]{Steven Ahlen}
\affiliation{Physics Dept., Boston University, 590 Commonwealth Avenue, Boston, MA 02215, USA}
\author[0000-0001-9712-0006]{Davide Bianchi}
\affiliation{Dipartimento di Fisica ``Aldo Pontremoli'', Universit\`a degli Studi di Milano, Via Celoria 16, I-20133 Milano, Italy}
\author{David Brooks}
\affiliation{Department of Physics \& Astronomy, University College London, Gower Street, London, WC1E 6BT, UK}
\author{Todd Claybaugh}
\affiliation{Lawrence Berkeley National Laboratory, 1 Cyclotron Road, Berkeley, CA 94720, USA}
\author[0000-0002-5954-7903]{Shaun Cole}
\affiliation{Institute for Computational Cosmology, Department of Physics, Durham University, South Road, Durham DH1 3LE, UK}
\author[0000-0002-1769-1640]{Axel de la Macorra}
\affiliation{Instituto de F\'{\i}sica, Universidad Nacional Aut\'{o}noma de M\'{e}xico,  Cd. de M\'{e}xico  C.P. 04510,  M\'{e}xico}
\author{Peter Doel}
\affiliation{Department of Physics \& Astronomy, University College London, Gower Street, London, WC1E 6BT, UK}
\author[0000-0002-3033-7312]{Andreu Font-Ribera}
\affiliation{Department of Physics \& Astronomy, University College London, Gower Street, London, WC1E 6BT, UK}\affiliation{Institut de F\'{i}sica d’Altes Energies (IFAE), The Barcelona Institute of Science and Technology, Campus UAB, 08193 Bellaterra Barcelona, Spain}
\author[0000-0002-2890-3725]{Jaime E. Forero-Romero}
\affiliation{Departamento de F\'isica, Universidad de los Andes, Cra. 1 No. 18A-10, Edificio Ip, CP 111711, Bogot\'a, Colombia}\affiliation{Observatorio Astron\'omico, Universidad de los Andes, Cra. 1 No. 18A-10, Edificio H, CP 111711 Bogot\'a, Colombia}
\author{Enrique Gaztañaga}
\affiliation{Institut d'Estudis Espacials de Catalunya (IEEC), 08034 Barcelona, Spain}\affiliation{Institute of Cosmology and Gravitation, University of Portsmouth, Dennis Sciama Building, Portsmouth, PO1 3FX, UK}\affiliation{Institute of Space Sciences, ICE-CSIC, Campus UAB, Carrer de Can Magrans s/n, 08913 Bellaterra, Barcelona, Spain}
\author[0000-0003-3142-233X]{Satya Gontcho A Gontcho}
\affiliation{Lawrence Berkeley National Laboratory, 1 Cyclotron Road, Berkeley, CA 94720, USA}
\author{Gaston Gutierrez}
\affiliation{Fermi National Accelerator Laboratory, PO Box 500, Batavia, IL 60510, USA}
\author{Klaus Honscheid}
\affiliation{Center for Cosmology and AstroParticle Physics, The Ohio State University, 191 West Woodruff Avenue, Columbus, OH 43210, USA}\affiliation{Department of Physics, The Ohio State University, 191 West Woodruff Avenue, Columbus, OH 43210, USA}\affiliation{The Ohio State University, Columbus, 43210 OH, USA}
\author[0000-0002-1081-9410]{Cullan Howlett}
\affiliation{School of Mathematics and Physics, University of Queensland, 4072, Australia}
\author{Stephanie Juneau}
\affiliation{NSF NOIRLab, 950 N. Cherry Ave., Tucson, AZ 85719, USA}
\author{Robert Kehoe}
\affiliation{Department of Physics, Southern Methodist University, 3215 Daniel Avenue, Dallas, TX 75275, USA}
\author[0000-0003-2644-135X]{Sergey Koposov}
\affiliation{Institute for Astronomy, University of Edinburgh, Royal Observatory, Blackford Hill, Edinburgh EH9 3HJ, UK}\affiliation{Institute of Astronomy, University of Cambridge, Madingley Road, Cambridge CB3 0HA, UK}
\author{Andrew Lambert}
\affiliation{Lawrence Berkeley National Laboratory, 1 Cyclotron Road, Berkeley, CA 94720, USA}
\author[0000-0003-1838-8528]{Martin Landriau}
\affiliation{Lawrence Berkeley National Laboratory, 1 Cyclotron Road, Berkeley, CA 94720, USA}
\author[0000-0001-7178-8868]{Laurent Le Guillou}
\affiliation{Sorbonne Universit\'{e}, CNRS/IN2P3, Laboratoire de Physique Nucl\'{e}aire et de Hautes Energies (LPNHE), FR-75005 Paris, France}
\author[0000-0002-1125-7384]{Aaron Meisner}
\affiliation{NSF NOIRLab, 950 N. Cherry Ave., Tucson, AZ 85719, USA}
\author{Ramon Miquel}
\affiliation{Instituci\'{o} Catalana de Recerca i Estudis Avan\c{c}ats, Passeig de Llu\'{\i}s Companys, 23, 08010 Barcelona, Spain}\affiliation{Institut de F\'{i}sica d’Altes Energies (IFAE), The Barcelona Institute of Science and Technology, Campus UAB, 08193 Bellaterra Barcelona, Spain}
\author[0000-0002-2733-4559]{John Moustakas}
\affiliation{Department of Physics and Astronomy, Siena College, 515 Loudon Road, Loudonville, NY 12211, USA}
\author[0000-0002-1544-8946]{Gustavo Niz}
\affiliation{Departamento de F\'{i}sica, Universidad de Guanajuato - DCI, C.P. 37150, Leon, Guanajuato, M\'{e}xico}\affiliation{Instituto Avanzado de Cosmolog\'{\i}a A.~C., San Marcos 11 - Atenas 202. Magdalena Contreras, 10720. Ciudad de M\'{e}xico, M\'{e}xico}
\author[0000-0002-0644-5727]{Will Percival}
\affiliation{Department of Physics and Astronomy, University of Waterloo, 200 University Ave W, Waterloo, ON N2L 3G1, Canada}\affiliation{Perimeter Institute for Theoretical Physics, 31 Caroline St. North, Waterloo, ON N2L 2Y5, Canada}\affiliation{Waterloo Centre for Astrophysics, University of Waterloo, 200 University Ave W, Waterloo, ON N2L 3G1, Canada}
\author[0000-0001-7145-8674]{Francisco Prada}
\affiliation{Instituto de Astrof\'{i}sica de Andaluc\'{i}a (CSIC), Glorieta de la Astronom\'{i}a, s/n, E-18008 Granada, Spain}
\author[0000-0001-6979-0125]{Ignasi Pérez-Ràfols}
\affiliation{Departament de F\'isica, EEBE, Universitat Polit\`ecnica de Catalunya, c/Eduard Maristany 10, 08930 Barcelona, Spain}
\author{Graziano Rossi}
\affiliation{Department of Physics and Astronomy, Sejong University, Seoul, 143-747, Korea}
\author[0000-0002-9646-8198]{Eusebio Sanchez}
\affiliation{CIEMAT, Avenida Complutense 40, E-28040 Madrid, Spain}
\author{Michael Schubnell}
\affiliation{Department of Physics, University of Michigan, Ann Arbor, MI 48109, USA}\affiliation{University of Michigan, Ann Arbor, MI 48109, USA}
\author[0000-0002-6588-3508]{Hee-Jong Seo}
\affiliation{Department of Physics \& Astronomy, Ohio University, Athens, OH 45701, USA}
\author{David Sprayberry}
\affiliation{NSF NOIRLab, 950 N. Cherry Ave., Tucson, AZ 85719, USA}
\author[0000-0003-1704-0781]{Gregory Tarlé}
\affiliation{University of Michigan, Ann Arbor, MI 48109, USA}
\author{Benjamin Alan Weaver}
\affiliation{NSF NOIRLab, 950 N. Cherry Ave., Tucson, AZ 85719, USA}
\author[0000-0002-6684-3997]{Hu Zou}
\affiliation{National Astronomical Observatories, Chinese Academy of Sciences, A20 Datun Rd., Chaoyang District, Beijing, 100012, P.R. China}

\begin{abstract}

We present three separate void catalogs created using a volume-limited sample of the DESI Data Release 1 Bright Galaxy Survey. We use the algorithms VoidFinder and V$^2$ to construct void catalogs out to a redshift of $z=0.24$. Excluding voids affected by the boundaries of the survey, we obtain 1489 voids with VoidFinder, 389 with V$^2$ using REVOLVER pruning, and 297 with V$^2$ using VIDE pruning. Comparing our catalogs with overlapping Sloan Digital Sky Survey void catalogs, we find generally consistent void properties but significant differences in the void volume overlap, which we attribute to differences in the galaxy selection and survey masks. These catalogs are suitable for studying the variation in galaxy properties with cosmic environment and for cosmological studies.

\end{abstract}

\section{Introduction} \label{sec:intro}

On large scales, the structure of the Universe is organized into a complex cosmic web consisting of dense clusters and filaments and underdense voids \citep{gregory1978, joeveer1978, Springel_2006, van_de_Weygaert_2014, Libeskind_2017}. This large-scale structure is the result of the gravitational collapse of matter following initial inhomogeneities in the primordial Universe \citep{Bernardeau_2002, Cautun_2014}. In the present era, cosmic voids occupy the majority of the Universe's volume, with sizes on the order of tens of megaparsecs \citep{VAN_DE_WEYGAERT_2011, jennings2013abundance, Douglass_2023}. Voids evolve under the combined effects of gravity, which expels their matter content, and dark energy, which dominates their expansion at an earlier time compared to surrounding dense regions \citep{Sheth_2004, Demchenko_2016}. In this way, voids are dynamically unique objects.

Voids provide a cosmological probe for models of dark energy and modified gravity \citep{van_de_Weygaert_2014, pisani2019cosmic}. The on-average spherical shape of voids makes them well suited for Alcock-Paczynski tests of cosmology \citep{1979Natur.281..358A, sutter2012first, Sutter_2014, PhysRevD.100.023504}. The theoretical distribution of void sizes, known as the void size function, is a probe of dark energy \citep{Sheth_2004,jennings2013abundance,Pisani_2015,Contarini_2022, contarini2023cosmological, song2024cosmological}. The void size function and void density profiles also serve as tests of modified gravity  \citep{PhysRevD.95.024018, perico2019cosmic}.

The constraining power of void-based cosmological tests benefits from large void catalogs. Recent years have seen void catalogs built using galaxies from the Sloan Digital Sky Survey \citep[SDSS;][]{Pan_2012, Sutter_2012, Nadathur_2016, Mao_2017, Achitouv_2019, Hamaus_2020, Douglass_2023}, with SDSS providing spectroscopy for millions of galaxies and quasars out to redshifts past $z>2$ \citep{Abazajian_2009,Abdurrouf2022}. Compared to SDSS, the Dark Energy Spectroscopic Instrument (DESI) Survey will provide a new benchmark for galaxy-redshift surveys, with measured redshifts for over 40 million targets out to redshifts past $z>3$ \citep{DESI2016a.Science,DESI2016b.Instr}. DESI uses four distinct spatially overlapping classes of galaxies to probe a large range of cosmic history. Compared to SDSS, void catalogs built with DESI will contain significantly higher void counts and will trace the evolution of voids over a larger interval of cosmic time.

Making full use of DESI galaxies in the construction of void catalogs requires consideration of the survey target selection. DESI tracers are selected using magnitude and color \citep{Hahn_2023, zhou2023,raichoor2023,chaussidon2023}, and the resulting change in tracer density with redshift and tracer type must be accounted for by void-finding algorithms. 

Previous void catalogs have accounted for the variation in tracer density with redshift by restricting their void finding to low-redshift, volume-limited galaxy samples \citep{Pan_2012,Douglass_2023}, or by weighting the performance of the void-finding algorithms with respect to the tracer density as a function of redshift \citep{Nadathur_2016, Mao_2017,Hamaus_2020}. While galaxy weighting has the advantages of obtaining a larger void sample and probing a greater range of cosmic history, it comes with the limitation of detecting artificially large voids due to the mean galaxy separation exceeding the expected void sizes \citep{Mao_2017}. Void catalogs built with SDSS at high redshift tend to find an abundance of voids with radii greater than 30 Mpc h$^{-1}$ \citep{Sutter_2012, Nadathur_2016}, whereas low-redshift, volume-limited catalogs tend to find an abundance of voids around the range of 15-20 Mpc h$^{-1}$ \citep{Pan_2012, Douglass_2023}, contrary to theoretical expectations. We expect void sizes to increase with decreasing redshift as we trace the evolution of expanding underdensities\footnote{Voids may contract over time when embedded within a larger, overdense region. This so-called ``void-in-cloud'' process affects the shape of the void size distribution, while the average void size increases with time \citep{van_de_Weygaert_2014}.} \citep{Sheth_2004}. That we instead see larger voids at higher redshifts may be indicative of the challenge of overcoming Malmquist bias when using galaxies to identify low-density regions in the cosmic web \citep{sutter2014sample}.

An accurate description of void sizes is important to astrophysical studies that observe the variation of galaxy properties with cosmic environment \citep{Patiri_2006, zaidouni2024impact}. In addition to high-redshift void catalogs, there is good reason for creating low-redshift, volume-limited catalogs, where we have the highest confidence in our ability to identify low-density voids.

In this paper, we present low-redshift void catalogs built on the DESI Bright Galaxy Survey (BGS). We use BGS Bright galaxies within $z\leq0.24$ to construct void catalogs with two void-finding algorithms: the sphere-growing algorithm VoidFinder \citep{El_Ad_1997,Hoyle_2002} and the watershed algorithm ZOBOV \citep{Neyrinck_2008}, titled V$^2$ in our implementation. 

Our void catalogs are collected in the DESIVAST value-added catalog. DESIVAST Data Release 1 (DR1) will be publicly released with DESI DR1. Our void catalogs will be updated with each data release; we here present results for DESI DR1. 

This paper is structured as follows. Section \ref{sec:algorithms} discusses the void-finding algorithms used for the void catalog construction. Section \ref{sec:data} discusses the creation of a volume-limited galaxy catalog from DESI BGS galaxies. Section \ref{sec:catalog} presents the void catalogs and compares the catalogs across different void-finding algorithms and to voids found in the common volume observed with SDSS. Section \ref{sec:conclusion} concludes and sets the scope for future iterations of DESIVAST. Acknowledgments are given in Section \ref{sec:thanks}, and an appendix with catalog file descriptions and data access instructions is presented in Section \ref{sec:appendix}.

\section{Void-finding Algorithms}\label{sec:algorithms}

We perform void finding with the sphere-growing algorithm VoidFinder \citep{El_Ad_1997,Hoyle_2002} and the watershed ZOBOV algorithm \citep{Neyrinck_2008}, here titled V$^2$. We use implementations of these two algorithms in the Void Analysis Software Toolkit\footnote{\url{https://github.com/DESI-UR/VAST}} \citep[VAST;][]{Douglass_2023}.

VoidFinder uses a volume-limited galaxy sample for void construction. The algorithm partitions the input galaxy sample into highly clustered ``wall'' regions and isolated ``field'' regions. Voids are defined to contain field galaxies and to be bordered by wall galaxies. The wall-field classification of galaxies is made based on the third-nearest neighbor of each galaxy, where galaxies that have a third-nearest neighbor above a threshold distance are identified as field galaxies. The threshold distance is taken to be $d = \bar{d} + 1.5\sigma$, where $\bar{d}$ is the average third-nearest neighbor separation and $\sigma$ is its standard deviation. The algorithm removes all field galaxies so that voids may be identified in the ``hole'' regions left behind. Next, the user defines a fine spatial grid in the survey volume with a resolution that is small relative to the expected sizes of the smallest voids. We use the default resolution of 5 Mpc h$^{-1}$. VoidFinder then seeds spheres in all empty grid cells, and the spheres are grown until they are bordered by four wall galaxies. The largest resulting spheres that do not overlap with any larger sphere are labeled as ``maximal spheres'' and are identified with individual voids. All remaining spheres that overlap 50\% of their volume with a maximal sphere are then merged with the maximal sphere, allowing for VoidFinder voids to describe the complex shapes expected from real voids, rather than modeling each void as a simple sphere \citep{El_Ad_1997,Hoyle_2002,Douglass_2023}.

In contrast to VoidFinder, V$^2$ may use a magnitude-limited galaxy sample. The algorithm uses the galaxy density field to define voids, which is calculated by applying a Voronoi tessellation to the galaxy catalog. The volume of a galaxy's Voronoi cell is used to estimate the local density. In the case of a magnitude-limited survey, the density for each cell can be weighted by the survey selection function. We have not applied such weights and therefore restrict ourselves to a volume-limited sample. The algorithm then identifies low-density basins of adjacent cells, termed ``zones'', each of which is separated by high-density ridges. Various combinations of zones may be defined as voids, and there exist differing treatments for combing zones, referred to as pruning methods. The VIDE pruning method \citep{sutter2014vide} uses a maximum linking density threshold between adjacent zones to determine which zones should be combined into voids and applies an optional central-density cut to the final catalog. The VIDE catalog consequently takes the form of a hierarchy of overlapping voids that consist of zones joined at various linking densities. The REVOLVER pruning method does not combine any zones and instead treats all individual zones as voids \citep{PhysRevD.100.023504}. The VAST software suite includes both pruning methods, referred to as V$_{\text{VIDE}}^2$ and V$_{\text{REVOLVER}}^2$. Any Voronoi cell extending outside of the survey mask is given an infinite density so as to prevent voids from exceeding the survey boundaries.

VoidFinder voids have been shown to correspond to dynamically distinct regions with low shell-crossing numbers \citep{veyrat2023voidfinding}. In comparison, V$^2$ voids are known to include wall features within the void volumes and combine adjacent dynamically distinct void regions into single voids. As result of these differences, VoidFinder voids tend to be smaller and more numerous than V$^2$ voids. VoidFinder voids have also been shown to contain galaxies that are bluer, fainter, and have higher specific star formation rates than the surrounding wall galaxies \citep{zaidouni2024impact}. V$^2$ voids in contrast contain galaxies with no notable differences from those in wall regions.

Thus, the watershed algorithms defined in V$^2$ and the sphere-growing algorithm VoidFinder each have distinct use cases. The watershed algorithms, which have relatively few tunable parameters and can be applied to magnitude-limited samples using weights accounting for the tracer selection function, are popular in cosmology. The VoidFinder algorithm, which must be optimized for tracer density and is challenging to apply beyond volume-limited samples, is preferable for studying the properties of galaxies as a function of their environment, due to its superior identification of dynamically distinct regions. Since both types of analyses are useful, we produce catalogs using both algorithms applied to a common volume-limited sample of DESI DR1 data.

\section{Data and Selection}\label{sec:data}

DESI is an ongoing, state-of-the-art galaxy-redshift survey that will cover 14,000 deg$^2$ of the sky and measure the redshifts of over 40 million galaxies and quasars out to z $>$ 3 \citep{Snowmass2013.Levi, DESI2016a.Science}. DESI has carried out the three-band Legacy Imaging Survey \citep{LS.Overview.Dey.2019}, from which targets have been selected for an ongoing 360–980 nm spectroscopic survey, conducted with the DESI instrument on the Mayall Telescope at Kitt Peak National Observatory \citep{DESI2016b.Instr, DESI2022.KP1.Instr, FocalPlane.Silber.2023, Corrector.Miller.2023}. DESI targets include overlapping populations of nearby bright galaxies, \citep{Hahn_2023}, luminous red galaxies \citep[LRGs;][]{zhou2023}, emission line galaxies \citep{raichoor2023}, and quasars \citep{chaussidon2023}. To facilitate a complex survey with multiple target classes, DESI has extensive data collection and processing pipelines, including procedures for survey tiling \citep{SurveyOps.Schlafly.2023}, spectroscopic reduction \citep{Spectro.Pipeline.Guy.2023}, and redshift fitting \citep{Redrock.Bailey.2024}.

DESI has completed survey validation \citep{DESI2023a.KP1.SV}, an early data release \citep{DESI2023b.KP1.EDR}, and DR1 data collection \citep{desi2024desiI}. DR1 science results include two point clustering measurements \citep{desi2024desiII}, baryon acoustic oscillations (BAO) from galaxies and quasars \citep{desicollaboration2024desi} and from the Lyman-alpha forest \citep{2024arXiv240403001D}, full-shape analysis from galaxies and quasars \citep{desi2024desiV}, and cosmological constraints from both BAO \citep{2024arXiv240403001D} and full-shape measurements \citep{desi2024desiVII}.

We construct a volume-limited catalog from the DESI DR1 BGS Bright survey \citep{Hahn_2023} as the galaxy input for our void finding. We select galaxies brighter than $M_r\leq-20$, a population that is known to trace large-scale structure \citep{Pan_2012}. 

We obtain K corrections for the galaxies from the FastSpecFit VACs \citep{fastspecfit23,moustakas:inprep}.
We select synthesized SDSS $r$-band absolute magnitudes ($M_r$) that are K corrected to $z=0.1$. We add an evolutionary (``E'') correction centered at the same redshift, $E(z)=-Q_0(z-0.1)$, with the intention of selecting galaxies at different redshifts that trace the same large-scale structure. The scale factor $Q_0$ is chosen as $Q_0=0.97$, following the calibration from \cite{10.1093/mnras/stu1886}. 

\begin{figure}
\centering
\includegraphics[width=.45\textwidth]{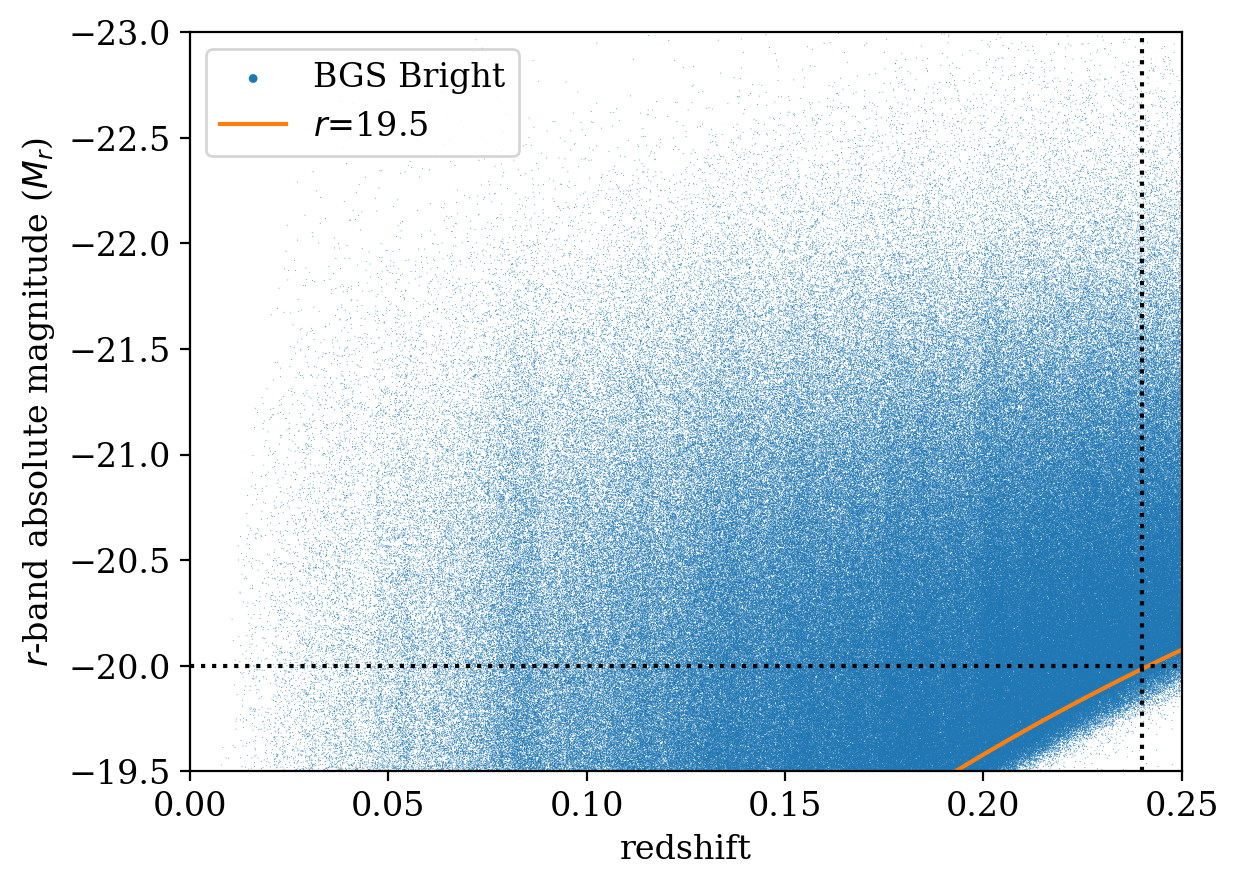} 
\caption{A magnitude-redshift diagram for BGS Bright galaxies, with K corrections and E corrections applied. The magnitude and redshift boundaries for the volume-limited sample are shown as dashed lines. The survey apparent magnitude limit is shown as an orange line.} 
\label{Fig.selection} 
\end{figure}

With K corrections and E corrections applied, our $M_r\leq-20$ mag cut corresponds to a redshift limit of $z\leq0.24$, the maximum distance at which $M_r=-20$ galaxies can be observed by the BGS Bright survey. This distance limit is a consequence of the $r\leq19.5$ apparent magnitude limit for BGS Bright. A magnitude-redshift plot of the galaxies and the selections for the volume-limited catalog are shown in Figure \ref{Fig.selection}. 

\begin{figure}
\centering
\includegraphics[width=.45\textwidth]{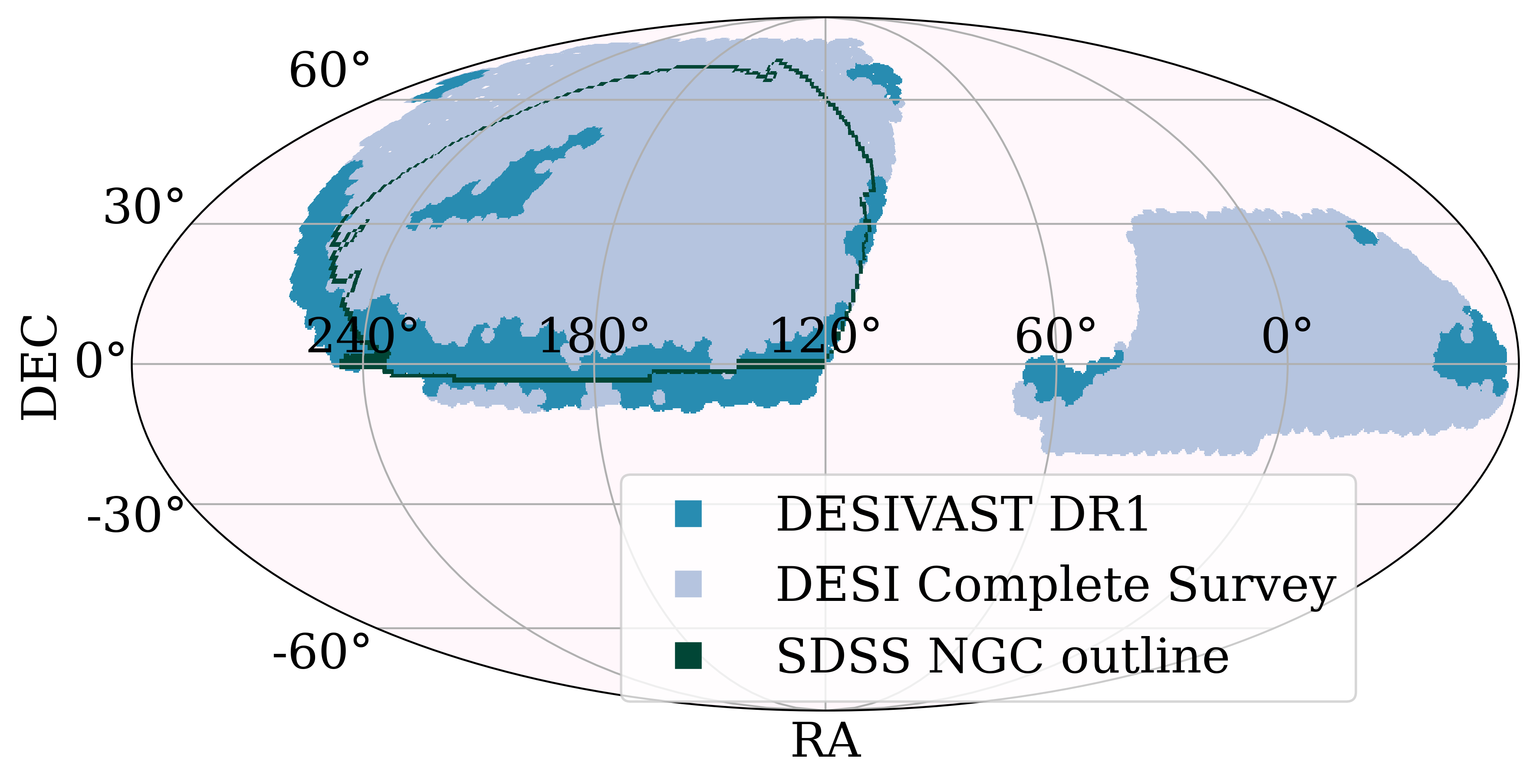} 
\caption{The angular mask used for the DESIVAST DR1 void catalogs, shown in teal. The complete DESI survey mask is shown in gray. For comparison, an outline of the SDSS NGC, which overlaps with the DESI NGC, is shown in dark green.} 
\label{Fig.mask} 
\end{figure}

The DR1 catalog has uneven angular coverage on the sky, with coverage consisting of one to four overlapping survey tiles. To minimize the effect of this variation, we impose an angular mask selecting only regions where all planned survey tiles for up to four passes have been observed. As DESI is an ongoing survey, this mask has many small, noncontiguous regions. We then use a smoothed version of the mask to select voids for our final catalogs. To smooth our mask, we convolve it with a Gaussian filter with $\sigma=2.8^\circ$. This convolution assigns each pixel a weight between zero and one. We keep mask pixels whose weights are greater than 0.34, finding that the resulting mask removes small, noncontiguous regions while preserving the survey edges. This final angular mask is shown in Figure \ref{Fig.mask} (labeled DESIVAST DR1) and covers a region of 2320 deg$^2$ in the north Galactic cap (NGC) and 386 deg$^2$ in the south Galactic cap (SGC). 

DESI redshifts are fitted using the Redrock software package \citep{Spectro.Pipeline.Guy.2023}. Redrock includes several fit quality indicators, notably a ZWARN flag that indicates a lack of confidence in the fit or errors in the fitting procedure, and a $\Delta\chi^2$ (DELTACHI2) indicator of how confidently the software has chosen the correct spectral classification for each spectral fit. To ensure accurate redshifts, we select only galaxies with no raised ZWARN flags and require $\Delta\chi^2>45$ for confident spectral classification. 

The angular mask and redshift cuts reduce the BGS DR1 sample from an initial count of 4,494,248 galaxies to a subsequent count of 1,208,667 galaxies. The $M_r\leq-20$ mag cut further reduces the sample to 484,412 galaxies. The spectroscopic quality cuts impose a minimal reduction in the galaxy sample size of 1.0\%. Our final volume-limited catalog contains 479,486 galaxies, with 400,668 in the NGC, and 78,818 in the SGC. The resulting survey volume is 0.085 Gpc$^3$.

\subsection{SDSS Data Release 7 Galaxies}\label{sec:data.sdss}

SDSS \citep{York_2000} was the most ambitious galaxy-redshift survey prior to DESI, conducted with the 2.5 m Sloan Foundation Telescope at the Apache Point Observatory \citep{Gunn_2006}. SDSS included a five-band photometric imaging survey \citep{Fukugita_1996} and a 380–920 nm spectroscopic survey \citep{Strauss_2002}, for overlapping populations of bright galaxies \citep{Strauss_2002}, LRGs \citep{Eisenstein_2001}, and quasars \citep{Richards_2002}. The seventh data release \citep[DR7;][]{Abazajian_2009} of SDSS presents the complete bright galaxy and LRG samples of the survey's first two stages, SDSS-I/II. These catalogs provide photometry for 357 million objects and 1.6 million spectra. Further data releases past SDSS DR7 do not provide additional high-tracer density, magnitude-complete samples, making SDSS DR7 the most recent galaxy-redshift survey suitable for a comparison with the DESIVAST sample.

SDSS DR7 covers a contiguous 7,500 deg$^2$ region of the NGC \citep{Abazajian_2009}, providing significant overlap with the complete DESI survey. We are interested in the consistency of voids found between the two surveys in their common region. Given 1,860 deg$^2$ overlap between the DESIVAST DR1 and SDSS NGC footprints, we are limited in DESIVAST DR1 by the number of SDSS voids that can be compared against DESI voids. However, it is still instructive to compare the two catalogs.

To enable comparisons between DESI and SDSS, we construct a volume-limited catalog of SDSS galaxies from the NASA-Sloan Atlas \citep[NSA;][]{Blanton_2011}. Version 1.0.1 of the NSA catalog provides spectroscopy for 641,409 galaxies in SDSS DR7. The SDSS DR7 void catalog of \cite{Douglass_2023} is built from the NSA galaxy catalog, K corrected to $z=0.0$ and with no applied E correction. For our own SDSS void catalogs, we chose to K correct the NSA catalog to $z=0.1$ and to apply an E correction $E(z)=-0.97(z-0.1)$, consistent with that used in DESIVAST.

An absolute magnitude cut of $M_r\leq-20$ imposes a $z\leq0.114$ redshift limit, the maximum distance at which $M_r=-20$ galaxies can be observed by the SDSS Main Galaxy Survey. We additionally impose an angular mask on the NSA catalog. The mask, shown in Figure \ref{Fig.mask}, selects galaxies within the SDSS NGC. 

Our angular mask and redshift cut select 379,405 galaxies from the NSA catalog. Our $M_r\leq-20$ mag cut selects a final population of 177,407 galaxies.

\subsection{The Impact of DESI and SDSS Galaxies on Voids}\label{sec:data.compare}

The completeness of SDSS and DESI affect their resulting void catalogs. The SDSS Main Galaxy Sample (MGS) is $>$99\% complete down to an $r$-band apparent magnitude of $r\leq17.77$, though 6\% of the sample lack spectroscopy due to fiber collisions \citep{Strauss_2002}. DESI BGS has $>$80\% fiber assignment completeness down to an $r$-band apparent magnitude of $r\leq19.5$ in its original four-pass program \citep{Hahn_2023}. Both DESI BGS and SDSS MGS have high redshift success rates such that almost all observed targets have successful redshifts.

When considering a volume-limited galaxy sample, we obtain fewer DESI galaxies than SDSS galaxies. This has the expected effect of producing larger voids in DESI than SDSS, as the wall regions bordering VoidFinder voids will lose galaxies and the watershed ridges bordering V$^2$ voids will lose resolution. 

To directly compare the volume-limited galaxy populations of the two surveys, we smooth the common DESIVAST-SDSS mask to 345 deg$^2$, thereby reducing edge effects. Our smoothing follows the same procedure used in section \ref{sec:data}, except we only keep mask pixels with weights greater than 0.99 in order to remove regions near the survey edges. In the resulting fiducial volume extending to $z=0.114$, we find that BGS galaxies have an average separation of 5.62 Mpc, while NSA galaxies have an average separation of 5.20 Mpc. 

\begin{figure}
\includegraphics[width=0.45\textwidth]{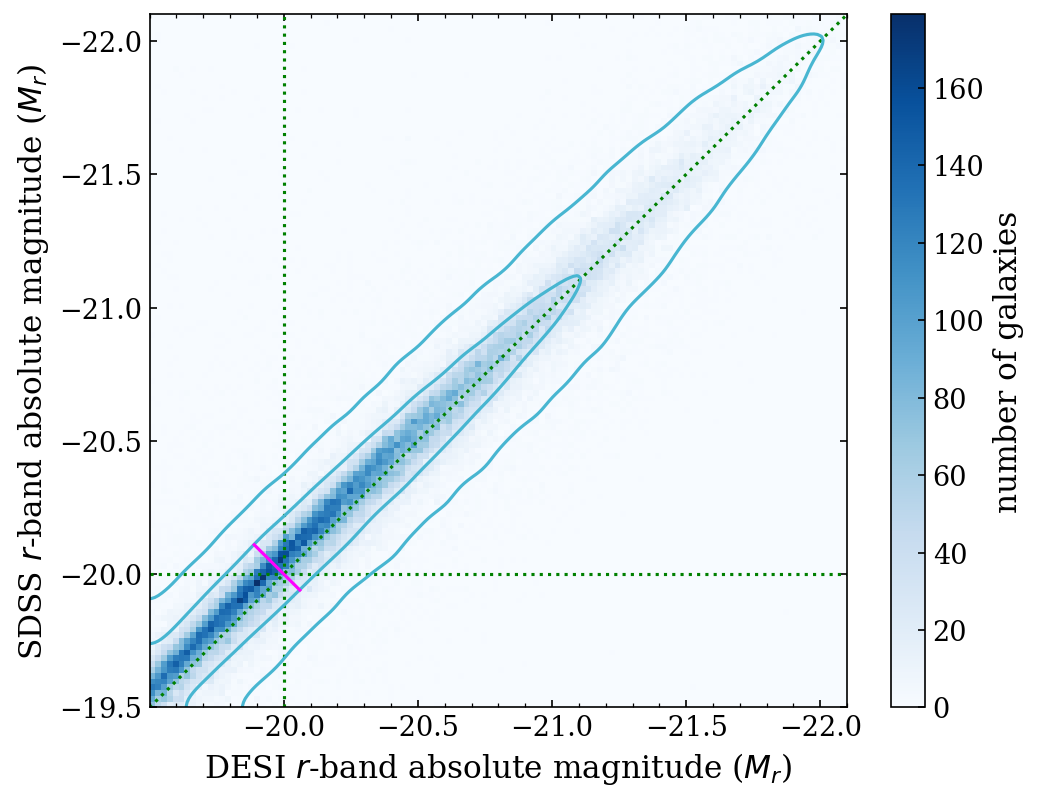} 
\caption{A two-dimensional histogram comparing the $r$-band magnitudes of galaxies reported by DESI and SDSS. The ideal of a one-to-one relation and the $M_r=-20$ mag thresholds are shown as dashed lines. The 68\% confidence interval at $M_r=-20$ is shown in magenta.} 
\label{Fig.kcorr} 
\end{figure}

The DESI and SDSS galaxy samples used for void finding are also affected by the galaxy magnitudes reported in each survey. In Figure \ref{Fig.kcorr}, we compare the magnitudes reported by the FastSpecFit VAC and the NSA catalog for matched galaxies. There is scatter in the agreement of the magnitudes, as well as a systematic offset, which affects which galaxies make the $M_r\leq-20$ cut for each survey and ultimately impacts the resulting void catalogs. The upper-left region of the plot, bounded by dashed green lines, shows galaxies included in SDSS but not DESI, and the lower-right region of the plot shows galaxies included in DESI but not SDSS. We see that at $M_r=-20$, galaxies tend to be reported as brighter in SDSS than in DESI. This has the effect of including more galaxies in the SDSS volume-limited catalog compared to DESI, once again suggesting that SDSS voids will be smaller than DESI voids.

We compare the galaxy membership of SDSS and DESI in our smoothed fiducial volume. We find that 30.80\% of the SDSS galaxies are not present in DESI. When accounting for galaxies that pass the $M_r\leq-20$ threshold for SDSS but not DESI, we find that 24.70\% of the SDSS galaxies are not present in DESI. Similarly, we find that 12.58\% of the DESI galaxies are not present in SDSS. Accounting for galaxies that pass the $M_r\leq-20$ threshold for DESI but not SDSS, we find that 10.82\% of the DESI galaxies are not present in SDSS.

\subsection{Mock Galaxy Catalogs}\label{sec:data.mocks}

To estimate the impact of statistical uncertainties and cosmic variance on the DESI void catalogs, we use simulated realizations of BGS galaxy catalogs. Specifically, we use Alternate Merged Target Ledger (AMTL) catalogs that simulate DESI fiber assignment \citep{lasker2024production} on mock catalogs \citep{desicollaboration2024desi} derived from the AbacusSummit N-body simulation suite \citep{10.1093/mnras/stab2484,10.1093/mnras/stab2482,10.1093/mnras/stab2980}. We select 25 AMTL catalogs with a cosmology using $\Omega_m = 0.315$, $\Omega_\Lambda = 0.685$, and $H_0=67.32 $ km s$^{-1}$ Mpc$^{-1}$ \citep{planckcollaboration2021planck}. The AMTL galaxies include $r$-band absolute magnitudes that allow us to apply the DESIVAST volume-limited catalog creation procedure to each mock. We take the spread in void properties across mocks to represent the effect of statistical uncertainties and cosmic variance on our void catalogs. 

To provide projections for a complete DESI survey BGS void catalog, we use a mock BGS galaxy catalog \citep{desicollaboration2024desi} derived from the AbacusSummit N-body simulation suite. We apply the complete DESI survey mask and the DESIVAST redshift limit to the mock catalog to obtain the desired survey volume.

To estimate the impact of statistical uncertainties and cosmic variance on SDSS voids, we use 25 galaxy mocks \citep{Hong_2016} built from the Horizon Run 4 N-body simulation \citep{Kim_2015}. The Horizon Run 4 mocks are built on a cosmology with $\Omega_m = 0.241$, $\Omega_\Lambda = 0.759$, and $H_0=71.9 $ km s$^{-1}$ Mpc$^{-1}$ \citep{Dunkley_2009}. The galaxy mocks include a mass estimate that serves as a proxy for luminosity. We apply the SDSS NGC angular mask, distance limits, and magnitude cut to each mock to create volume-limited catalogs. The spread in the resulting void properties is taken to represent the effect of statistical uncertainties and cosmic variance.

As the Horizon Run 4 and Abacus AMTL galaxy mocks use different cosmologies, we limit their use to comparing the variance, and not the expected value, of void catalog properties.

\section{The DESIVAST DR1 Void Catalogs}
\label{sec:catalog}

We construct three DESI void catalogs, one each for VoidFinder, V$_{\text{VIDE}}^2$, and V$_{\text{REVOLVER}}^2$ voids. We assume a cosmology with $\Omega_m = 0.315$, $\Omega_\Lambda = 0.685$, and $H_0=67.32 $ km s$^{-1}$ Mpc$^{-1}$ \citep{planckcollaboration2021planck} to convert galaxy redshifts to comoving distances. For all catalogs, we impose a minimum radius cut to remove voids that are more likely to be Poisson-fluctuated underdensities in the galaxy distribution (see \cite{Hoyle_2002} for details). In practice, we require VoidFinder maximal spheres to have a radius larger than 10 Mpc h$^{-1}$, following a cut optimized for SDSS by \cite{Pan_2012}. Given our similar target density in BGS at the redshifts being investigated, we apply the same cut to the DESIVAST DR1 sample. Similarly, we cut V$^2$ voids whose effective spherical radii are smaller than 10 Mpc h$^{-1}$.

\begin{figure}
\includegraphics[width=0.45\textwidth]{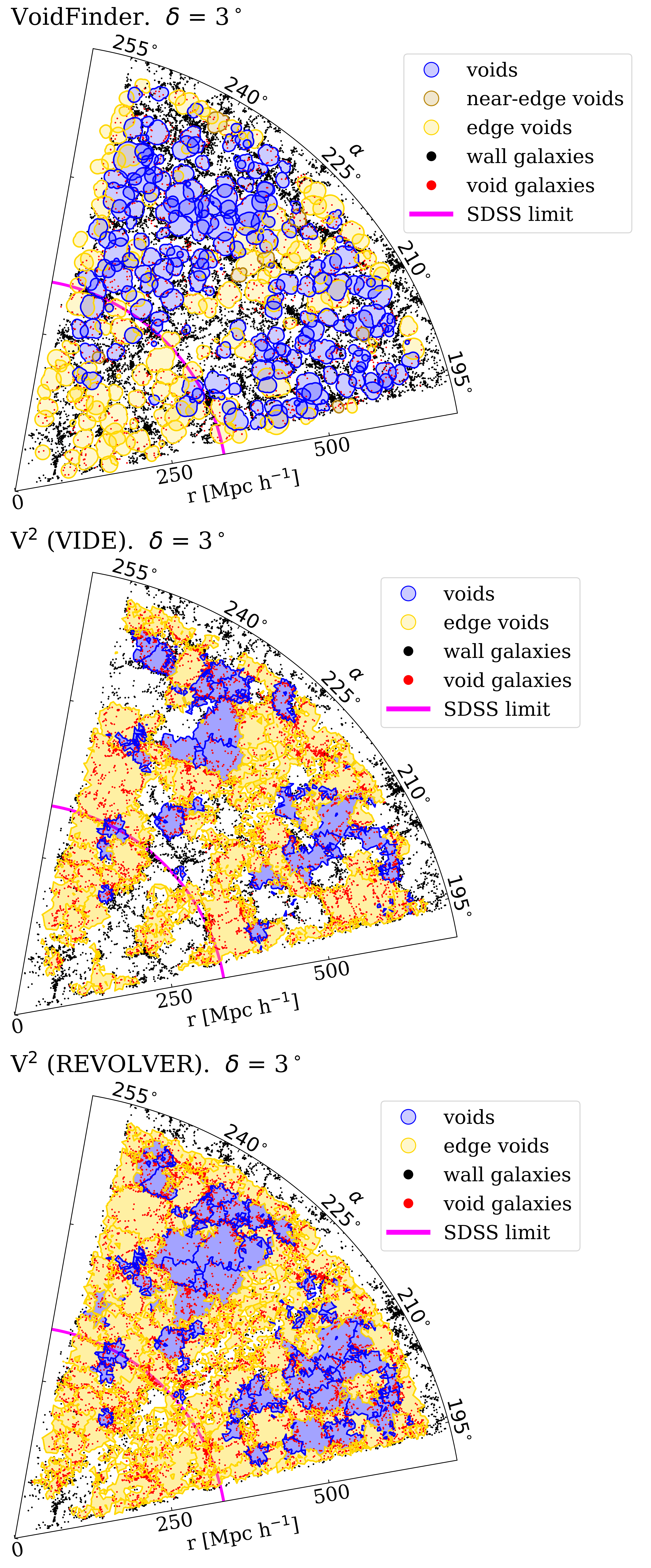} 
\caption{Slice plots of the three void catalogs (from top to bottom: VoidFinder, V$^2_{\text{VIDE}}$, and V$^2_{\text{REVOLVER}}$) overlaid on the same survey region. A 10 Mpc h$^{-1}$ thick slice of galaxies is plotted over the void distribution at a decl. of $\delta=3^\circ$. Void (wall) galaxies are shown in red (black). Edge (interior) voids are shown in yellow (blue). We show a comparison with the redshift limit of the \cite{Douglass_2023} SDSS void catalogs (solid magenta line).} 
\label{Fig.sliceplots} 
\end{figure}

The physical distribution of voids resulting from each algorithm is shown in Figure \ref{Fig.sliceplots}, displayed in a cross section of the catalogs at a decl. of 3$^\circ$. Voids that fall near the survey boundaries, termed ``edge voids'', are drawn in yellow. The remaining interior voids are drawn in blue. We also show galaxies that fall within 5 Mpc h$^{-1}$ of the cross-sectional plane, rendered in black (red) when they fall exterior (interior) to voids. Since the DESIVAST DR1 footprint is constrained to a narrow strip of decl. near the celestial equator, the effect of the survey edge is apparent in Figure \ref{Fig.sliceplots}, where voids in several bands of R.A. and at the redshift limit of the sample are classified as touching the edges of the survey. For comparison, we show the distance limit of the \cite{Douglass_2023} SDSS void catalogs, demonstrating the improved depth of redshift coverage with DESI.

Qualitatively, VoidFinder voids appear smaller and more numerous than V$^2$ voids, and V$^2$ voids appear more likely to include wall galaxies within voids, particularly for V$^2_{\text{REVOLVER}}$. This behavior is consistent with the known properties of voids for each algorithm; VoidFinder voids enclose dynamically distinct regions, while V$^2$ voids combine multiple VoidFinder voids into single objects \citep{Douglass_2023,veyrat2023voidfinding}.

V$^2_{\text{VIDE}}$ voids appear to fill less of the survey volume than V$^2_{\text{REVOLVER}}$ voids due to the central-density cut that is (optionally) part of V$^2_{\text{VIDE}}$. Following the convention in \cite{sutter2014vide}, we remove V$^2_{\text{VIDE}}$ void candidates with a central density that is higher than the V$^2_{\text{VIDE}}$ minimum zone-linking density. V$^2$, being a watershed algorithm, is designed to fill 100\% of the survey volume with voids. The nonvoid regions seen in both V$^2$ catalogs are due to our central-density and minimum-size cuts.

Edge voids occur because our void-finding algorithms define voids based on the galaxy distribution. The galaxy survey boundaries impact the accurate detection of voids, distorting the shapes of voids in their vicinity, making it useful to identify voids near the survey edge. We define edge voids for VoidFinder as those whose volumes exceed the survey boundaries. For VoidFinder alone, we additionally identify a class of so-called ``near-edge voids'' that are contained by the survey but have at least one hole falling within 10 Mpc h$^{-1}$ of the survey boundaries. For V$^2$, voids do not exceed survey boundaries by construction, and we identify edge voids as those with over $10\%$ of their surface area touching the survey boundary. This definition separates two distinct populations of voids: those with minimal contact with the survey boundaries, and those with significant contact with the survey boundaries. \cite{Douglass_2023} showed that this edge void definition accurately identifies V$^2$ voids whose volumes are impacted by the survey boundaries.

For comparison to DESI voids, we construct three SDSS void catalogs, one each for VoidFinder, V$_{\text{VIDE}}^2$, and V$_{\text{REVOLVER}}^2$. These catalogs are equivalent to those in \cite{Douglass_2023}, but with the void and wall galaxies having K corrections and E corrections consistent with DESIVAST.

\subsection{Comparison of Void Catalogs}
\label{sec:catalog.comp}

To compare the sizes of voids across algorithms, we define an effective void radius as the radius corresponding to a sphere with the same volume as the overall void. The void volume is calculated with Monte Carlo sampling of a bounding region around the void. The use of an effective void radius allows us to compare void sizes across algorithms that define voids differently.

The void counts and catalog properties for all three void catalogs are presented in Tables \ref{tab:comp} and \ref{tab:vol}. We compare the median effective radii, maximum effective radii, volumes belonging to voids, and galaxies belonging to voids for each algorithm. We also include comparisons to the SDSS void catalogs from \cite{Douglass_2023}, with K corrections and E corrections described in Section \ref{sec:data.sdss}. 

We report uncertainties in the void catalog properties due to cosmic variance and the magnitude systematic uncertainties of different surveys. We take the standard deviation of each catalog property across 25 AMTL mock void catalogs $\sigma_c$ to represent the combined effects of statistical uncertainty and cosmic variance for DESI. We do the same with 25 Horizon Run 4 mock void catalogs for SDSS.

To evaluate the systematic uncertainty due to magnitude, we take the width of the 68\% confidence interval of the scatter at $M_r=-20$ in Figure \ref{Fig.kcorr} to represent the range over which the void-finding algorithm outputs vary due to the systematic uncertainties in the magnitudes of DESI and SDSS. We create versions of the DESIVAST and SDSS VAST catalogs with alternate magnitude cuts at the edges of the 68\% interval range projected onto the axes of Figure \ref{Fig.kcorr}: $M_r\leq-19.89$ and $M_r\leq-20.06$ for DESI, and $M_r\leq-19.94$ and $M_r\leq-20.11$ for SDSS. The differences in the void catalog properties between each alternate magnitude threshold and the main void catalogs produce asymmetric uncertainties $\sigma^{+}_{-}$ in the DESIVAST and SDSS VAST catalogs due to the scatter in the reported magnitudes.

For a given void catalog property, we report the uncertainty on a value $x$ in as $x \pm \sigma_c$(stat.)$^{+\sigma^{+}}_{-\sigma_{-}}$(sys.). In some instances, the values $x_1$ and $x_2$ for the alternate magnitude cut catalogs are both greater or lesser than the main catalog value $x$. In this case, we take the greater of the $\sigma^{+}_{-}$ uncertainties to represent one of the asymmetric uncertainties and report the other as zero.

When reporting median effective void radii, we account for the standard error of the median $\sigma_{\tilde{x}}$ as part of the statistical uncertainty. We report the uncertainty on a median value $\tilde{x}$ as $\tilde{x} \pm (\sigma_s)$(stat.)$^{+\sigma^{+}}_{-\sigma_{-}}$(sys.) where $\sigma^2_s = \sigma^2_c + \sigma^2_{\tilde{x}}$.

In Table \ref{tab:comp}, we present the void counts for DESIVAST DR1, the projected void counts for the complete DESI survey, and the median and maximum effective void radii. For DESIVAST DR1, we find 1,489 interior voids with VoidFinder, 389 with V$^2_{\text{REVOLVER}}$, and 297 with V$^2_{\text{VIDE}}$. The V$^2_{\text{VIDE}}$ algorithm is unique in that it identifies a hierarchy of overlapping voids. We note that of the 297 (1,478) interior (total) voids in the V$^2_{\text{VIDE}}$ catalog, a count of 249 (1,257) are top-level ``parent'' voids in the hierarchy that do not overlap with one another.

Across all algorithms, over 60\% of the detected voids are edge voids. This is due to the thin angular mask used to construct the DESIVAST DR1 catalogs, which can be seen in Figure \ref{Fig.mask}. We compare void counts with our SDSS void catalogs, where we find that despite the thin angular coverage of the DESIVAST DR1 mask, the DESIVAST void counts surpass or are comparable to the corresponding SDSS counts due to DESI's improved redshift coverage, which allows for a distance limit in the volume-limited galaxy catalog of 677 Mpc h$^{-1}$, compared with the SDSS limit of 332 Mpc h$^{-1}$.

\begin{figure*}
\centering
\includegraphics[width=1.\textwidth]{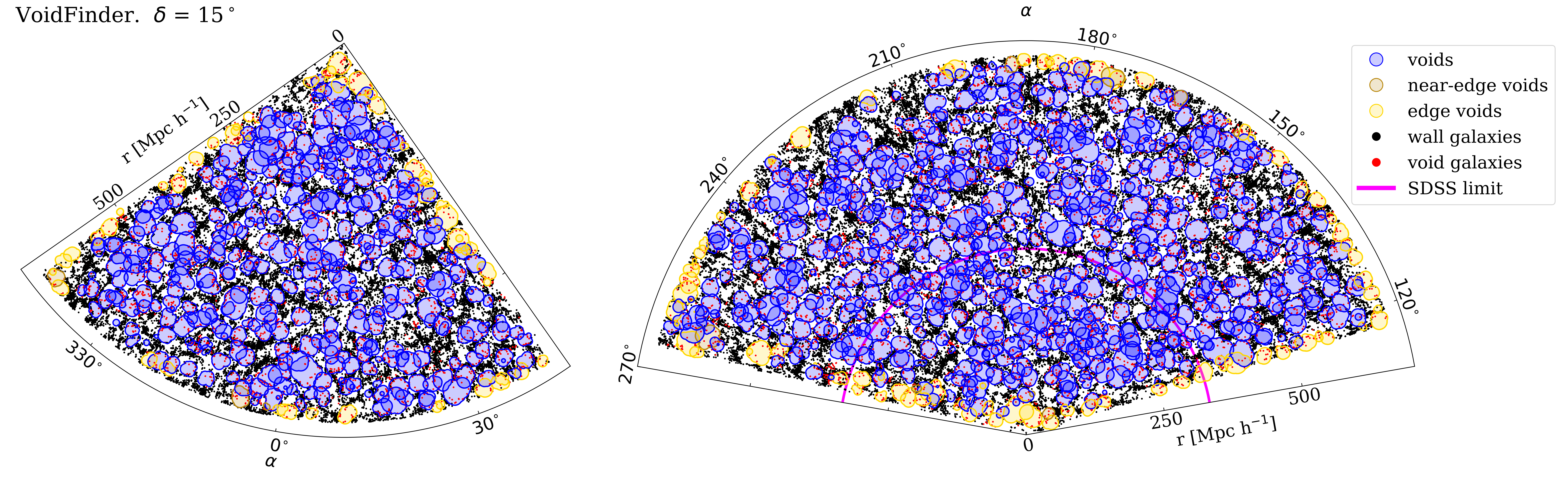} 
\caption{Slice plots of the projected VoidFinder volume-limited mock void catalog for the complete DESI survey. The catalog is built on an AbacusSummit BGS mock. A 10 Mpc h$^{-1}$ thick slice of mock galaxies is plotted over the void distribution at a decl. of $\delta=15^\circ$. Void (wall) galaxies are shown in red (black). Edge (interior) voids are shown in yellow (blue). The SGC is shown on the left, and the NGC is shown on the right. For the NGC, we show a comparison with the \cite{Douglass_2023} SDSS void catalog redshift limit (solid magenta line).} 
\label{Fig.mock} 
\end{figure*}

We use a BGS mock galaxy catalog to project the number of voids expected in the equivalent complete DESI survey version of our volume-limited catalog. We find around 15,000 interior voids in the VoidFinder catalog, and around 7,000 interior voids in each V$^2$ catalog. In Figure \ref{Fig.mock}, we show the physical distribution of voids in the complete DESI survey VoidFinder mock catalog. In contrast to the DESIVAST DR1 catalogs in Figure \ref{Fig.sliceplots}, we see that the complete DESI survey mock catalogs occupy the majority of the survey volume with a contiguous region of interior voids. This reflects the change in the relative number of edge and interior voids between DR1 and the complete DESI survey.

\begin{deluxetable*}{lccc}
  \tablewidth{0pt}
  \tablecaption{DESI Void Catalog Comparisons\label{tab:comp}}
    \tablehead{\colhead{Catalog} & \colhead{VoidFinder} & \colhead{\Vsquared/REVOLVER} & \colhead{\Vsquared/VIDE}}
    \startdata
    DESIVAST Void count (total/interior) & 3,765 / 1,489 & 1,992 / 389 & 1,478 / 297
    \\
    DESIVAST NGC Void Count (total/interior) & 3,241 / 1,333 & 1,692 / 355 & 1,258 / 269
    \\
    DESIVAST SGC Void Count (total/interior) & 524 / 156 & 300 / 27 & 223 / 27
    \\
    DESIVAST NGC Void Count in $V_{\text{fid}}$ (total/interior) & 929 / 859 & 550 / 300 & 401 / 225
    \\
    DESIVAST SGC Void Count in $V_{\text{fid}}$ (total/interior) & 93 / 83 & 67 / 27 & 53 / 21
    \\
    DESIVAST Complete Projected Void Count (total/interior) & 18,181 / 15,385 & 8,635 / 6,789 & 9,648 / 7,290
    \\
    SDSS DR7 Void Count (total/interior) & 1,187 / 810 & 684 / 374 & 500 / 275
    \\
    \hline
    \multirow{2}{*}{Median $R_{\text{eff}}$ in $V_{\text{fid}}$ [Mpc h$^{-1}$] (DESI/SDSS)} & $15.7\pm0.1^{+0.0}_{-0.1}$ / & $19.8\pm 0.2^{+0.01}_{-0.8}$ / & $19.3\pm0.3^{+0.0}_{-1.0}$ / \\
     & $15.9 \pm0.1^{+0.1}_{-0.2}$ & $18.8\pm0.4^{+0.0}_{-0.4}$ &  $18.0\pm0.5^{+0.2}_{-0.0}$
    \\
    \hline
    Maximum $R_{\text{eff}}$ in $V_{\text{fid}}$ [Mpc h$^{-1}$] (DESI/SDSS) & 31.7 / 30.2 & 43.5 / 42.1 & 55.9 / 56.5
    \\
  \enddata
  \tablecomments{Properties of the DESIVAST DR1 void catalogs for VoidFinder, V$^2_{\text{REVOLVER}}$, and V$^2_{\text{VIDE}}$, with a fiducial volume $V_{\text{fid}}$ used to account for survey-edge effects. Projections for the completed DESI survey are given with AbacusSummit BGS mocks. Our SDSS catalogs serve as a modification to the work of \cite{Douglass_2023}. For the median radii $\tilde{x}$, we report statistical uncertainties  $\sigma_s$, and asymmetric systematic uncertainties from galaxy magnitudes, $\sigma_+$ and $\sigma_-$, written as $\tilde{x}\pm {\sigma_s}^{+\sigma_+}_{-\sigma_-}$.}
\end{deluxetable*}

\begin{deluxetable*}{lcccc}
  \tablewidth{0pt}
  \tablecaption{Void-finding Algorithm Comparisons\label{tab:vol}}
  \tablehead{\colhead{Algorithm} & \colhead{\% $V_{\text{fid}}$ in Voids} & \colhead{\% $V_{\text{fid}}$ in Voids} & \colhead{\% Galaxies in Voids} & \colhead{\% Galaxies in Voids} \\
  \colhead{} & \colhead{(DESI DR1)} & \colhead{(SDSS DR7)} & \colhead{(DESI DR1)} & \colhead{(SDSS DR7)}}
    \startdata
    VoidFinder & 
    $62.8\pm1.2^{+1.3}_{-2.1}$ & 
    $62.7\pm0.8^{+2.1}_{-1.0}$ & 
    $22.8\pm0.8^{+1.2}_{-2.0}$ & 
    $20.3\pm0.5^{+1.9}_{-0.9}$
    \\
    \Vsquared/REVOLVER & 
    $98.7\pm0.2^{+0.0}_{-0.2}$ & 
    $99.0\pm0.4^{+0.3}_{-0.0}$ & 
    $98.6\pm0.3^{+0.0}_{-0.5}$ & 
    $98.3\pm0.6^{+0.5}_{-0.8}$ 
    \\
    \Vsquared/VIDE & 
    $69.3\pm3.0^{+3.0}_{-0.0}$ &
    $69.9\pm3.2^{+3.8}_{-0.0}$ & 
    $69.7\pm2.6^{+2.7}_{-0.0}$ & 
    $69.6\pm3.2^{+3.4}_{-0.0}$ 
    \\
    \hline
    VoidFinder $\cap$ \Vsquared/REVOLVER & 
    $62.2\pm1.2^{+1.3}_{-2.1}$ & 
    $62.5\pm0.8^{+2.0}_{-1.0}$ & 
    $22.6\pm0.8^{+1.1}_{-2.0}$ & 
    $20.3\pm0.5^{+1.9}_{-1.0}$
    \\ 
    VoidFinder $\cap$ \Vsquared/VIDE & 
    $43.6\pm2.0^{+1.2}_{-0.0}$ & 
    $44.0\pm1.9^{+3.8}_{-0.7}$ & 
    $16.1\pm0.8^{+1.0}_{-0.9}$ & 
    $14.3\pm0.7^{+2.1}_{-0.6}$ 
    \\
    \Vsquared/VIDE $\cap$ \Vsquared/REVOLVER & 
    $69.3\pm3.0^{+3.0}_{-0.0}$ & 
    $69.9\pm3.2^{+3.7}_{-0.0}$ & 
    $69.7\pm2.6^{+2.7}_{-0.0}$ & 
    $69.6\pm3.1^{+3.4}_{-0.0}$
  \enddata
  \tablecomments{The void volume fractions and fraction of galaxies in voids for each void-finding algorithm, with a fiducial volume $V_{\text{fid}}$ used to account for survey-edge effects. We also show the agreement between each of the different algorithms. Our SDSS catalogs serve as a modification to the work of \cite{Douglass_2023}. For each value $x$, we report uncertainties due to cosmic variance, $\sigma_c$, and asymmetric uncertainties from galaxy magnitudes, $\sigma_+$ and $\sigma_-$, written as $x\pm {\sigma_c}^{+\sigma_+}_{-\sigma_-}$.}
\end{deluxetable*}

To reduce the effect of the survey boundaries on the statistical properties of voids, we report void properties within an interior fiducial volume $V_{\text{fid}}$. We exclude any region within 30 Mpc h$^{-1}$ of the survey boundaries, removing regions near the DESIVAST DR1 mask edges and redshift limits. We likewise define a fiducial volume $V_{\text{fid}}$ for SDSS by removing regions within 30 Mpc h$^{-1}$ of the SDSS survey boundaries. Table \ref{tab:comp} shows that the fiducial volume significantly reduces the number of edge voids that impact the reported void statistics, while allowing for the reported statistics to be calculated across a common volume for each algorithm.

The median and maximum effective radii for each algorithm of voids within our fiducial volume $V_{\text{fid}}$ are presented in Table \ref{tab:comp}. The median effective radii are 15.7$\pm0.1$(stat.)$^{+0.0}_{-0.1}$(sys.) Mpc h$^{-1}$ for VoidFinder, 19.8$\pm0.2$(stat.)$^{+0.01}_{-0.8}$(sys.) Mpc h$^{-1}$ for V$^2_{\text{REVOLVER}}$, and 19.3$\pm0.3$(stat.)$^{+0.0}_{-1.0}$(sys.) Mpc h$^{-1}$ for V$^2_{\text{VIDE}}$. The maximum effective radius is 31.7 (43.5, 55.9) Mpc h$^{-1}$ for VoidFinder (V$^2_{\text{REVOLVER}}$, V$^2_{\text{VIDE}}$). These trends reflect how the V$^2$ algorithm discovers larger voids than VoidFinder. For each algorithm, the median effective radii of the DESI and SDSS catalogs agree within their uncertainties.

In Table \ref{tab:vol}, we compare the void volume fraction, meaning the fraction of the survey volume in voids, and the fraction of galaxies in voids across catalogs. We again use our fiducial volume $V_{\text{fid}}$ for this comparison. The void volume fractions are calculated by populating the fiducial volume with a grid of points spaced 1 Mpc h$^{-1}$ apart and counting the relative number of grid points located within and exterior to voids. We find that DESIVAST voids fill the majority of the volume, with 62.8$\pm 1.2$(stat.)$^{+1.3}_{-2.1}$(sys.)\% of the volume in VoidFinder Voids, 98.7$\pm 0.2$(stat.)$^{+0.0}_{-0.2}$(sys.)\% of the volume in V$^2_{\text{REVOLVER}}$ voids, and 69.3$\pm 3.0$(stat.)$^{+3.0}_{-0.0}$(sys.) \% of the volume in V$^2_{\text{VIDE}}$ voids. V$^2$ voids fill more of the Universe than VoidFinder voids, and V$^2_{\text{REVOLVER}}$ identifies the near entirety of the volume as belonging to voids. These differences follow from the volume-filling behavior of the V$^2$ algorithms and the postprocessing cuts that we have applied to the catalogs. The SDSS void volume fractions agree with the DESI void volume fractions within the uncertainties for all algorithms.

We select galaxies falling within our fiducial volume $V_{\text{fid}}$ to compare the galaxy membership in voids across algorithms. We find that 22.8$\pm 0.8$(stat.)$^{+1.2}_{-2.0}$(sys.)\% of galaxies are in VoidFinder voids, that 98.6$\pm 0.3$(stat.)$^{+0.0}_{-0.5}$(sys.)\% of galaxies are V$^2_{\text{REVOLVER}}$ voids, and that 69.7$\pm 2.6$(stat.)$^{+2.7}_{-0.0}$(sys.)\% of galaxies are V$^2_{\text{VIDE}}$ voids. The V$^2$ catalogs place a majority of the galaxies in voids, with V$^2_{\text{REVOLVER}}$ in particular identifying almost the entire galaxy population as belonging to voids. The high percentage of galaxies in V$^2$ voids are consequent to the volume-filling behavior of the V$^2$ algorithm and the postprocessing cuts that we have applied to the catalogs. The fraction of galaxies within SDSS voids is consistent with the fraction of galaxies within DESI voids across all algorithms. 

To further compare the different algorithms, we show in Table \ref{tab:vol} the void volume common to VoidFinder and V$^2_{\text{VIDE}}$, the void volume common to VoidFinder and V$^2_{\text{REVOLVER}}$, and the void volume common to V$^2_{\text{REVOLVER}}$ and V$^2_{\text{VIDE}}$. As before, we use the fiducial volume $V_{\text{fid}}$ for this comparison. We find that compared to the 62.8$\pm 1.2$(stat.)$^{+1.3}_{-2.1}$(sys.)\% of the fiducial volume within VoidFinder voids, a volume of  62.2$\pm 1.2$(stat.)$^{+1.3}_{-2.1}$(sys.)\% is also within V$^2_{\text{REVOLVER}}$ voids, and a volume of 43.6$\pm 2.0$(stat.)$^{+1.2}_{-0.0}$(sys.)\% is also within V$^2_{\text{VIDE}}$ voids. We find the entirety of the volume within V$^2_{\text{VIDE}}$ voids to be also within V$^2_{\text{REVOLVER}}$ voids. We also compare the agreement in galaxy membership within voids. We find that compared to the 22.8$\pm 0.8$(stat.)$^{+1.2}_{-2.0}$(sys.)\% of galaxies within VoidFinder voids, a sample of 22.6$\pm 0.8$(stat.)$^{+1.1}_{-2.0}$(sys.)\% is also within V$^2_{\text{REVOLVER}}$ voids, and a sample of 16.1$\pm 0.8$(stat.)$^{+1.0}_{-0.9}$(sys.)\% is also within V$^2_{\text{VIDE}}$ voids. As with the volume comparison, we find the entirety of the galaxies within V$^2_{\text{VIDE}}$ voids to belong also to V$^2_{\text{REVOLVER}}$ voids. The results for SDSS are in agreement with those for DESI across all algorithms.

We observe overall consistency in the properties of DESIVAST and SDSS voids. We find that our SDSS void catalog properties at times differ from those reported in \cite{Douglass_2023}. This occurs for three reasons. First, our modified K corrections and E corrections for the NSA catalog change the void-finding output. Second, \cite{Douglass_2023} place a 50\% median void radius cut on their V$^2_{\text{REVOLVER}}$ catalog, whereas we use a 10 Mpc h$^{-1}$ void radius cut. Third, we use a 30 Mpc h$^{-1}$ border for defining the SDSS fiducial volume when calculating certain catalog properties. This border is used to prevent survey-edge effects from impacting the void catalog properties. Such a border was not used in \cite{Douglass_2023}.

\subsection{The Volume Overlap of DESI and SDSS voids}
\label{sec:catalog.overlap}

\begin{deluxetable}{cccc}

\tablewidth{0pt}
 \tablecaption{Percentage of Fiducial Volume in Void Unions\label{tab:volcom}}
 \tablehead{\colhead{Survey Region} & \colhead{VoidFinder} & \colhead{V$^2_{\text{REVOLVER}}$} & \colhead{V$^2_{\text{VIDE}}$} \\ \colhead{} & \colhead{(\%)} & \colhead{(\%)} & \colhead{(\%)}}
 \startdata 
 DESI void $\cap$ SDSS void & 50.1 & 81.6 &  35.8\\
 DESI void $\cap$ SDSS nonvoid & 10.5 & 5.2 & 24.9\\
 DESI nonvoid $\cap$ SDSS void & 11.4 & 12.1 & 30.8\\
 DESI nonvoid $\cap$ SDSS nonvoid & 28.0 & 1.1 & 8.5\\
 \enddata
 \tablecomments{The percentage of the common DESI-SDSS fiducial volume found in the unions of void catalogs, for our three void-finding algorithms. The first row gives the percentage of the fiducial volume within the union of DESI and SDSS voids. The second row gives the percentage of the fiducial volume found in DESI voids but not in SDSS voids. The third row gives the percentage of the fiducial volume found in SDSS voids but not in DESI voids. The fourth row gives the percentage of the fiducial volume exterior to both SDSS and DESI voids.}
\end{deluxetable}

\begin{figure}
\centering
\includegraphics[width=.45\textwidth]{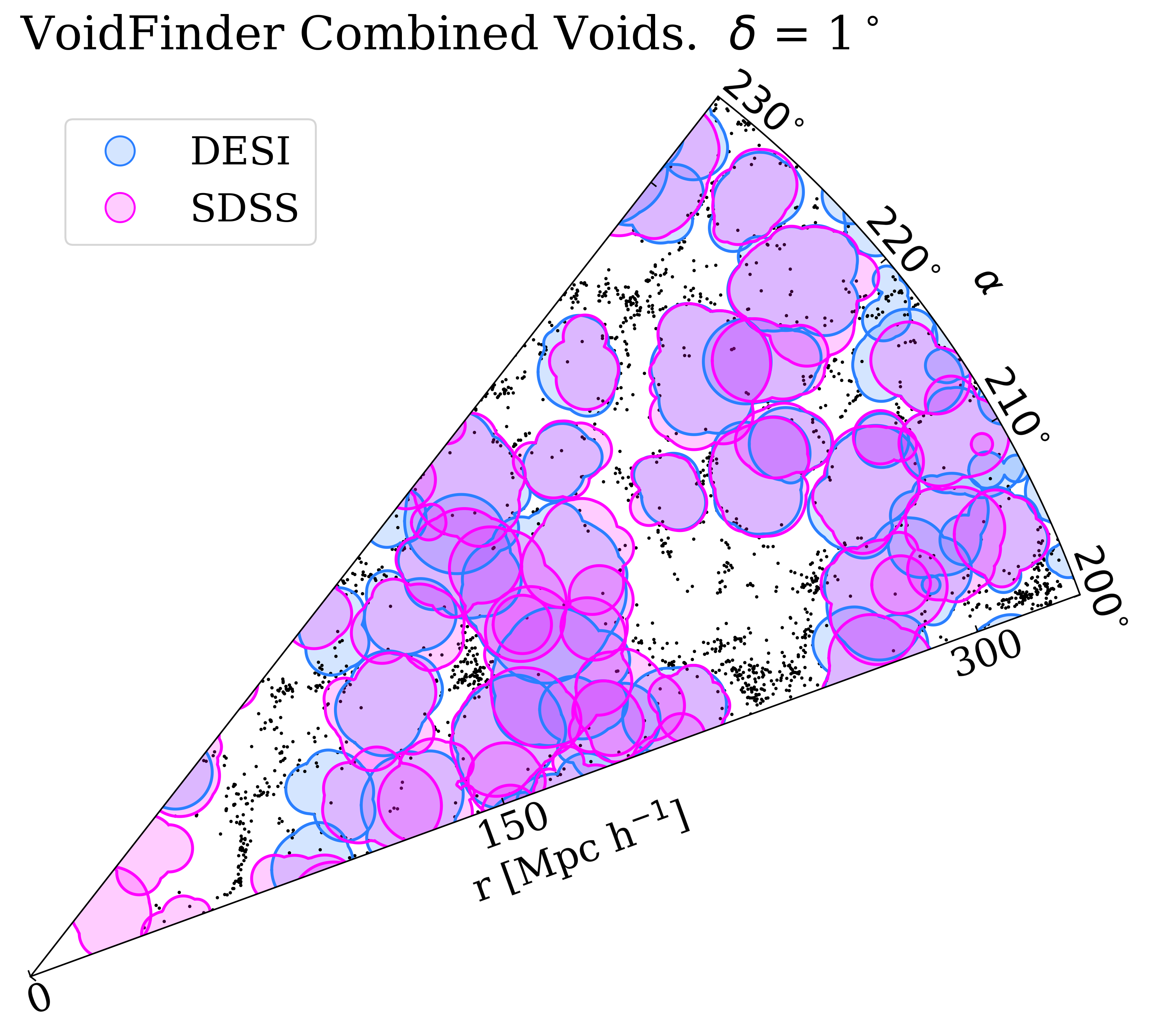} 
\caption{A slice plot of the VoidFinder void catalogs for DESI and SDSS. DESI voids are shown in cyan and SDSS voids are shown in magenta. A 10 Mpc h$^{-1}$ thick slice of DESI galaxies is plotted over the void distribution in black at a decl. of $\delta=1^\circ$.} 
\label{Fig.compare} 
\end{figure}

We examine the consistency of our DESIVAST and SDSS void catalogs. We consider voids located within the 1308 deg$^2$ angular mask representing the DESIVAST-SDSS footprint overlap and within the SDSS $z=0.114$ redshift limit. To reduce survey-edge effects, we apply a cut that removes regions within 10 Mpc h$^{-1}$ of the survey boundaries. We use a 10 Mpc h$^{-1}$ cut rather than the 30 Mpc h$^{-1}$ cut used for previous fiducial volumes, as the latter cut results in a vanishingly small number of voids. With our reduced edge cut, we expect DESI and SDSS survey-edge effects to have a greater effect on the reported void statistics than with our previous comparisons.

We show the volume overlap of DESIVAST and SDSS voids that fall within the fiducial region in Table \ref{tab:volcom}. For VoidFinder, we find that 50.1\% of the targeted region is contained within both DESIVAST and SDSS voids, whereas 10.5\% of the region is uniquely within DESI voids and 11.4\% of the region is uniquely within SDSS voids. The remaining region is exterior to both void catalogs. We emphasize that this indicates an overall agreement in the performance of VoidFinder between DESI and SDSS, with 81-83\% of the void volume in each catalog also belonging to voids in the other catalog. However, the remaining difference is significant, as we would expect a near 100\% agreement in the ideal case.

Slice plots of the DESI and SDSS void catalogs in the targeted region are shown in Figure \ref{Fig.compare}. We see that despite the 10\% level difference in the volume overlap of the catalogs, they identify the same general regions as voids.

For V$^2_{\text{VIDE}}$, we find that 35.8\% of the region is contained within both DESIVAST and SDSS voids. 24.9\% of the region is uniquely in DESIVAST voids, and 30.8\% of the region is uniquely within SDSS voids. 54-59\% of the void volume in each catalog then also belongs to voids in the other catalog. For V$^2_{\text{REVOLVER}}$, we find that 81.6\% of the region is contained within both DESIVAST and SDSS voids. 5.2\% of the region is uniquely in DESIVAST voids, and 12.1\% of the region is uniquely within SDSS voids. 87-94\% of the void volume in each catalog then also belongs to voids in the other catalog. When using a 30 Mpc h$^{-1}$ survey-edge cut on the fiducial volume, the V$^2_{\text{REVOLVER}}$ void catalog is volume filling, and we expect a near 100\% agreement for V$^2_{\text{REVOLVER}}$ between DESI and SDSS. When using a 10 Mpc h$^{-1}$ survey-edge cut, the V$^2_{\text{REVOLVER}}$ pruning output contains more small voids that do not pass our 10 Mpc h$^{-1}$ minimum void radius cut for the final catalog, resulting in greater disagreement for  V$^2_{\text{REVOLVER}}$ between DESI and SDSS.

The differences in the volume overlap of the DESIVAST and SDSS void catalogs can be attributed in part to differing survey completeness. Our SDSS volume-limited catalog has more galaxies than our DESI volume-limited catalog, which changes the third-nearest neighbor threshold used by VoidFinder and alters which galaxies are removed by the algorithm to grow holes in. Despite these changes, VoidFinder is able to identify the same general low-density regions as voids. 

A higher survey completeness causes V$^2$ to better resolve watershed ridges, leading to a larger number of zones prior to pruning. For VIDE, better-resolved watershed ridges also affect which adjacent zones meet the zone-linking criteria. This serves to change the quantity, size, and location of void candidates, which in turn affects which void candidates pass the central-density cut that we apply to the VIDE catalog. We see a severe change in which regions of the survey volume VIDE identifies as voids for our different galaxy samples.

The differences in the volume overlap of the void catalogs may also be explained by DESIVAST and SDSS having different angular masks. Nearly all DESIVAST voids within $z\leq0.114$ are edge voids. While most SDSS voids are not edge voids, the SDSS NGC mask defines a substantial part of the boundary of the DESIVAST-SDSS overlap footprint, thus introducing SDSS edge voids into the overlapping region. Both surveys then contribute edge voids into the shared fiducial volume. As the shapes of edge voids are uniquely impacted by their respective survey boundaries, a comparison of edge voids may contribute to the overall disagreement in the DESIVAST-SDSS void volume overlap.

We expect that the complete DESI survey, having an angular mask that encompasses the SDSS catalog, will provide a comparison of a larger number of interior voids between the two surveys. The survey angular masks will then no longer be as significant of a factor impacting the volume comparison.

We overall find that VoidFinder is more resilient than V$^2_{\text{VIDE}}$ to changes in survey design when it comes to consistently identifying the same regions as voids. We find as expected that V$^2_{\text{REVOLVER}}$ consistently fills the majority of both survey volumes with voids.

\begin{deluxetable*}{ccccccccccc}
  \tablewidth{0pt}
  \tablecaption{VoidFinder Maximal Spheres Header Data Unit Columns\label{tab:VFmax}}
    \tablehead{\colhead{X} & \colhead{Y} & \colhead{Z} & \colhead{RADIUS} & \colhead{VOID} & \colhead{EDGE} & \colhead{R} & \colhead{RA} & \colhead{DEC} & \colhead{R\_EFF} & \colhead{R\_EFF\_UNCERT} \\ \colhead{Mpc h$^{-1}$} & \colhead{Mpc h$^{-1}$} & \colhead{Mpc h$^{-1}$} & \colhead{Mpc h$^{-1}$} & \colhead{} & \colhead{} & \colhead{Mpc h$^{-1}$} & \colhead{deg} & \colhead{deg} & \colhead{Mpc h$^{-1}$} & \colhead{Mpc h$^{-1}$} \\  \colhead{(1)} & \colhead{(2)} & \colhead{(3)} & \colhead{(4)} & \colhead{(5)} & \colhead{(6)} & \colhead{(7)} & \colhead{(8)} & \colhead{(9)} & \colhead{(10)} & \colhead{(11)}}
    \startdata
    -624.51 & -133.79 & -85.41 & 24.04 & 0 & 1 & 644.37 & 192.09 & -7.62 & 30.44 & 0.108
    \\
    -224.19 & -355.12 & 59.92 & 23.98 & 1 & 1 & 424.22 & 237.74 & 8.12 & 31.69 & 0.114
    \\
    -140.92 & -585.75 & 152.59 & 23.51 & 2 & 1 & 621.49 & 256.47 & 14.21 & 30.08 & 0.108
  \enddata
  \tablecomments{The column organization of the VoidFinder maximal spheres header data unit (HDU), with each row corresponding to an example maximal sphere. Columns include (1--3) the Cartesian coordinates of the sphere's center, (4) the sphere's radius, (5) a flag identifying each sphere's void, (6) a Boolean edge-flag indicating if a void is an edge void, (7--9) the spherical coordinates of the sphere's center, and (10--11) the void's effective radius and its uncertainty. We show the first three voids in our VoidFinder catalog. The full catalog is published online at \url{https://data.desi.lbl.gov/public/dr1/vac/dr1/desivast/}.}
\end{deluxetable*}

\begin{deluxetable*}{ccccccccccc}
  \tablewidth{0pt}
  \tablecaption{V$^2$ Voids Header Data Unit Columns\label{tab:V2void}}
    \tablehead{\colhead{VOID} & \colhead{X} & \colhead{Y} & \colhead{Z} & \colhead{REDSHIFT} & \colhead{RA} & \colhead{DEC} & \colhead{RADIUS} & \colhead{X1\ldots} & \colhead{TOT\_AREA} & \colhead{EDGE\_AREA} \\ \colhead{} & \colhead{Mpc h$^{-1}$} & \colhead{Mpc h$^{-1}$} & \colhead{Mpc h$^{-1}$} & \colhead{} & \colhead{deg} & \colhead{deg} & \colhead{Mpc h$^{-1}$} & \colhead{Mpc h$^{-1}$} & \colhead{(Mpc h$^{-1}$)$^2$} & \colhead{(Mpc h$^{-1}$)$^2$} \\  \colhead{(1)} & \colhead{(2)} & \colhead{(3)} & \colhead{(4)} & \colhead{(5)} & \colhead{(6)} & \colhead{(7)} & \colhead{(8)} & \colhead{(9-17)} & \colhead{(18)} & \colhead{(19)}}
    \startdata
    0 & -217.74 & 442.78 & 238.31 & 0.192 & 116.19 & 25.78 & 29.32 & -4.13\ldots & 41568.33 & 14649.29
    \\
    1 & -36.36 & -269.64 & 85.79 & 0.0974 & 262.32 & 17.5 & 25.1 & -5.35\ldots & 27470.71 & 8740.29
    \\
    2 & -203.35 & -562.47 & 72.98 & 0.212 & 250.12 & 6.96 & 43.46 & -8.33\ldots & 76671.18 & 4926.86
  \enddata
  \tablecomments{The column organization of the V$^2$ voids HDU, with each row corresponding to an example void. Columns include (1) a flag identifying each void, (2--4) the Cartesian coordinates of the void's center, (5--7) the spherical coordinates of the void's center, (8) the void's effective radius, (9--17) the three components [X--Z] of the three ellipsoid axes [1--3] of the best-fit ellipsoid, (18) the void's surface area, and (19) the void's surface area touching the survey boundary. We show the first three voids in our V$^2_{\text{VIDE}}$ catalog. The full catalog is published online at \url{https://data.desi.lbl.gov/public/dr1/vac/dr1/desivast/}.}
\end{deluxetable*}

\begin{deluxetable}{ccccc}
  \tablewidth{0pt}
  \tablecaption{VoidFinder Holes Header Data Unit Columns\label{tab:VFhole}}
  
    \tablehead{\colhead{X} & \colhead{Y} & \colhead{Z} & \colhead{RADIUS} & \colhead{VOID} \\ 
    \colhead{Mpc h$^{-1}$} & \colhead{Mpc h$^{-1}$} & \colhead{Mpc h$^{-1}$} & \colhead{Mpc h$^{-1}$} & \colhead{}
    \\
    \colhead{(1)} & \colhead{(2)} & \colhead{(3)} & \colhead{(4)} & \colhead{(5)}}
  \startdata
    -624.51 & -133.79 & -85.41 & 24.04 & 0 \\
    -224.19 & -355.12 & 59.92 & 23.98 & 1 \\
    -224.23 & -355.06 & 59.8 & 23.97 & 1
  \enddata
  \tablecomments{The column organization of the VoidFinder hole spheres HDU, with each row corresponding to an example hole. Columns include (1--3) the Cartesian coordinates of the sphere's center, (4) the sphere's radius, and (5) a flag identifying each sphere's void.  We show the first three holes in our VoidFinder catalog. The full catalog is published online at \url{https://data.desi.lbl.gov/public/dr1/vac/dr1/desivast/}.}
\end{deluxetable}

\begin{deluxetable}{ccc}
  \tablewidth{0pt}
  \tablecaption{V$^2$ Zones Header Data Unit Columns\label{tab:V2zone}}
  
    \tablehead{\colhead{ZONE} & \colhead{VOID0} & \colhead{VOID1} \\ \colhead{\phantom{.}\hspace{.5cm}(1)\hspace{.5cm}\phantom{.}} & \colhead{\phantom{.}\hspace{.5cm}(2)\hspace{.5cm}\phantom{.}} & \colhead{\phantom{.}\hspace{.5cm}(3)\hspace{.5cm}\phantom{.}}}
  \startdata
    0 & 1257 & 1257 \\
    1 & -1 & -1 \\
    2 & -1 & -1 
  \enddata
  \tablecomments{The column organization of the V$^2$ zones HDU, with each row corresponding to an example zone. Columns include (1) the zone ID, (2) the ID of the zone's smallest containing void, and (3) the ID of the zone's largest containing void. Values of -1 indicate that a zone does not belong to any void.  We show the first three zones in our V$^2_{\text{VIDE}}$ catalog. The full catalog is published online at \url{https://data.desi.lbl.gov/public/dr1/vac/dr1/desivast/}.}
\end{deluxetable}

\begin{deluxetable*}{ccccccccc}
  \tablewidth{0pt}
  \tablecaption{V$^2$ Galaxies Header Data Unit Columns\label{tab:V2galzone}}
    \tablehead{ \colhead{GAL}  & \colhead{TARGET} & \colhead{X} & \colhead{Y} & \colhead{Z} & \colhead{ZONE} & \colhead{DEPTH} & \colhead{EDGE} & \colhead{OUT} \\ \colhead{} & \colhead{} & \colhead{Mpc h$^{-1}$} & \colhead{Mpc h$^{-1}$} & \colhead{Mpc h$^{-1}$} & \colhead{} & \colhead{} & \colhead{} & \colhead{} \\ \colhead{(1)}  & \colhead{(2)} & \colhead{(3)} & \colhead{(4)} & \colhead{(5)} & \colhead{(6)} & \colhead{(7)} & \colhead{(8)} & \colhead{(9)}}
   \startdata
    0 & 39627540897202342 & -549.38 & 210.35 & -105.09 & 2949 & 0 & 1 & 0 \\
    1 & 39627540897202561 & -313.07 & 119.55 & -60.12 & 2949 & 0 & 1 & 0 \\
    2 & 39627540897202679 & -521.83 & 198.87 & -99.74 & 2949 & 0 & 1 & 0
  \enddata
  \tablecomments{The column organization of the V$^2$ galaxies HDU, with each row corresponding to an example galaxy. Columns include (1) the V$^2$-specified galaxy ID, (2) the original survey-specified galaxy ID, (3--5) the Cartesian coordinates of the galaxy, (6) the ID of the zone containing the galaxy, (7) the minimum number of adjacent cells between the galaxy and the edge of its zone, (8) a Boolean flag indicating if the galaxy's cell is fully contained by the survey mask, and (9) a Boolean flag indicating if the galaxy is within the survey mask.  We show the first three galaxies in our V$^2_{\text{VIDE}}$ catalog. The full catalog is published online at \url{https://data.desi.lbl.gov/public/dr1/vac/dr1/desivast/}.}
\end{deluxetable*}

\begin{deluxetable}{ccc}
  \tablewidth{0pt}
  \tablecaption{V$^2$ Galaxy Visualization Header Data Unit Columns\label{tab:V2galviz}}
    \tablehead{\colhead{GID} & \colhead{G2V} & \colhead{G2V2} \\ \colhead{\phantom{.}\hspace{.6cm}1\hspace{.6cm}\phantom{.}} & \colhead{\phantom{.}\hspace{.6cm}2\hspace{.6cm}\phantom{.}} & \colhead{\phantom{.}\hspace{.6cm}3\hspace{.6cm}\phantom{.}}}
    \startdata
     0 & -1 & -1\\
     1 & -1 & -1\\
     2 & -1 & -1
  \enddata
  \tablecomments{The column organization of the V$^2$ galaxy visualization HDU, with each row corresponding to an example galaxy. Columns include (1) the galaxy ID, (2) the ID of the galaxy's containing void, and (3) the ID of the void containing the galaxy's nearest neighbor galaxy. Values of -1 indicate that a galaxy does not belong to a void.  We show the first three galaxies in our V$^2_{\text{VIDE}}$ catalog. The full catalog is published online at \url{https://data.desi.lbl.gov/public/dr1/vac/dr1/desivast/}.}
\end{deluxetable}

\begin{deluxetable}{ccc}
  \tablewidth{0pt}
  \tablecaption{V$^2$ Triangles Header Data Unit Columns\label{tab:V2tri}}
    \tablehead{\colhead{VOID\_ID} & \colhead{N\_X\ldots} & \colhead{P1\_X\ldots} \\
    \colhead{} & \colhead{Mpc h$^{-1}$} & \colhead{Mpc h$^{-1}$}
    \\
    \colhead{\phantom{.}\hspace{.5cm}1\hspace{.5cm}\phantom{.}} & \colhead{\phantom{.}\hspace{.5cm}(2--4)\hspace{.5cm}\phantom{.}} & \colhead{\phantom{.}\hspace{.5cm}(5--13)\hspace{.5cm}\phantom{.}}}
    \startdata 
     1 & 0.838\ldots & -29.11\ldots\\
     1 & -0.497\ldots & -33.54\ldots\\
     1 & -0.497\ldots & -33.54\ldots
  \enddata
  \tablecomments{The column organization of the V$^2$ triangles HDU, with each row corresponding to a triangle. Columns include (1) the ID of the void containing the triangle, (2--4) the three components [X--Z] of the triangle's normal vector, (5--13) the three components [X--Z] of the triangle's three vertices [1-3].  We show the first three triangles in our V$^2_{\text{VIDE}}$ catalog. The full catalog is published online at \url{https://data.desi.lbl.gov/public/dr1/vac/dr1/desivast/}.}
\end{deluxetable}

\section{Conclusion}\label{sec:conclusion}
We create three void catalogs over a volume-limited sample of DESI DR1 BGS Bright tracers within a redshift of $z\leq0.24$. Our catalogs are built from two different algorithms: the sphere-growing algorithm VoidFinder and the watershed algorithm V$^2$ with the VIDE and REVOLVER pruning methods. As a result of the thin survey mask of DESI DR1, our catalogs consist mainly of edge voids, which occur along the survey boundaries. We find 1,489 nonedge voids in the VoidFinder catalog, 389 in the V$^2_{\text{REVOLVER}}$ catalog, and 297 in the V$^2_{\text{VIDE}}$ catalog. Our VoidFinder voids are smaller, more numerous, and occupy less of the survey volume than our V$^2$ voids, consistent with the well-understood behavior of each algorithm. We find that void catalog properties such as void volume fractions and the fraction of galaxies within voids are consistent between DESI and SDSS voids.

We find that the volume overlap agreement between DESI and SDSS has significant differences for all algorithms, particularly for V$^2_{\text{VIDE}}$. We propose differences in the DESI and SDSS survey design as the reason for this disagreement. We note that despite the disagreement in the volume overlap, VoidFinder is able to identify the same general low-density regions as voids.

We present our DESI-SDSS void comparison as a precursor to the complete DESI survey, where we will be able to compare a larger, contiguous region of both surveys. Our future catalogs will also extend beyond a volume-limited BGS Bright sample to include the full set of DESI tracers, taking advantage of the wide range of cosmic history that DESI probes.

\section{Acknowledgements}\label{sec:thanks}

We would like to thank Michael Vogeley for his expertise and for useful discussions regarding void finding and void analysis. We would also like to thank Steve O'Neill for writing a significant portion of the VAST VoidFinder codebase. We express our gratitude for the helpful feedback provided by the anonymous referee.

We acknowledge support from the United States Department of Energy grant DE-SC0008475,
Experimental Studies of Elementary Particles and Fields for H.R., S.B and D.V. We acknowledge support from grant 62177 from the John Templeton Foundation for H.R, K.D., and D.V. 

This material is based upon work supported by the U.S. Department of Energy (DOE), Office of Science, Office of High-Energy Physics, under contract No. DE–AC02–05CH11231, and by the National Energy Research Scientific Computing Center, a DOE Office of Science User Facility under the same contract. Additional support for DESI was provided by the U.S. National Science Foundation (NSF), Division of Astronomical Sciences under Contract No. AST-0950945 to the NSF’s National Optical-Infrared Astronomy Research Laboratory; the Science and Technology Facilities Council of the United Kingdom; the Gordon and Betty Moore Foundation; the Heising-Simons Foundation; the French Alternative Energies and Atomic Energy Commission (CEA); the National Council of Humanities, Science and Technology of Mexico (CONAHCYT); the Ministry of Science, Innovation and Universities of Spain (MICIU/AEI/10.13039/501100011033), and by the DESI Member Institutions: \url{https://www.desi.lbl.gov/collaborating-institutions}. Any opinions, findings, and conclusions or recommendations expressed in this material are those of the author(s) and do not necessarily reflect the views of the U. S. National Science Foundation, the U. S. Department of Energy, or any of the listed funding agencies.

The authors are honored to be permitted to conduct scientific research on Iolkam Du’ag (Kitt Peak), a mountain with particular significance to the Tohono O’odham Nation.

Funding for the SDSS and SDSS-II has been provided by the Alfred P. Sloan Foundation, the Participating Institutions, the National Science Foundation, the U.S. Department of Energy, the National Aeronautics and Space Administration, the Japanese Monbukagakusho, the Max Planck Society, and the Higher Education Funding Council for England.  The SDSS website is \url{http://www.sdss.org/}.

The SDSS is managed by the Astrophysical Research Consortium for the Participating Institutions. The Participating Institutions are the American Museum of Natural History, Astrophysical Institute Potsdam, University of Basel, University of Cambridge, Case Western Reserve University, University of Chicago, Drexel University, Fermilab, the Institute for Advanced Study, the Japan Participation Group, Johns Hopkins University, the Joint Institute for Nuclear Astrophysics, the Kavli Institute for Particle Astrophysics and Cosmology, the Korean Scientist Group, the Chinese Academy of Sciences (LAMOST), Los Alamos National Laboratory, the Max Planck Institute for Astronomy (MPIA), the Max Planck Institute for Astrophysics (MPA), New Mexico State University, Ohio State University, University of Pittsburgh, University of Portsmouth, Princeton University, the United States Naval Observatory, and the University of Washington.

\section{Appendix}
\label{sec:appendix}

\subsection{Void Catalog File Descriptions for DESIVAST}
\label{sec:appendix.files}

Our three void catalogs will be included in DESIVAST, a VAC to be released with DESI DR1. In DESIVAST, the void catalogs are stored as FITS files and are separated into NGC and SGC catalogs. The principal file organization identifying unique voids for each algorithm is presented in Table \ref{tab:VFmax} for VoidFinder and Table \ref{tab:V2void} for V$^2$. 

For VoidFinder, the void information is separated into two FITS HDUs: a maximal spheres HDU and a holes HDU. The maximal spheres HDU is exemplified in Table \ref{tab:VFmax}. The HDU contains the location of the maximal spheres in Cartesian and spherical coordinates, the radius of the maximal spheres, a flag identifying the void each maximal sphere belongs to, an edge flag determining if each maximal sphere belongs to an edge or interior void, the effective radius of the void that each maximal sphere belongs to, and the uncertainty in the effective radius calculation. The holes HDU is exemplified in Table \ref{tab:VFhole}. The HDU contains the Cartesian coordinates of the holes, the hole radii, and a flag identifying with void each hole belongs to.

The V$^2$ catalog information is divided among five HDUs. Among these, the voids HDU, exemplified in Table \ref{tab:V2void}, includes the center of voids as the volume-weighted sum of their Voronoi cells:

\begin{equation}
\vec{x}_{\text{center}}=V_{\text{void}}^{-1}\sum_i\vec{x}_iV_i
\end{equation}

Here $V_{\text{void}}$ is the total void volume and $i$ sums over the cells to access their positions and volumes. The voids HDU contains a flag identifying each void, the weighted centers of each void in spherical and Cartesian coordinates, the effective radii of the voids, the axes of ellipsoids fitted to each void, the surface area of each void, and the partial surface area of each void that is adjacent to the survey boundaries. The zones HDU, exemplified in Table \ref{tab:V2zone}, contains a flag uniquely identifying each zone as well as the smallest and largest voids in the void hierarchy that each zone belongs to. The galaxies HDU, exemplified in Table \ref{tab:V2galzone}, contains a flag uniquely identifying each galaxy, the galaxy ID assigned by the original survey, the Cartesian coordinates of the galaxy, a flag for which zone each galaxy belongs to, the number of cells between a galaxy's cell and the edge of its zone, a flag for whether the galaxy's cell extends past the survey boundaries, and a flag for whether the galaxy itself is outside the survey boundaries. The galaxy visualization HDU, exemplified in Table \ref{tab:V2galviz}, contains a flag uniquely identifying each galaxy, a flag for which void each galaxy belongs to, and a flag for which void each galaxy's nearest neighbor galaxy belongs to. The triangles HDU, exemplified in Table \ref{tab:V2tri}, describes the triangles that may be used to visualize the voids in representative figures. The HDU columns include a flag for which void each triangle belongs to, three columns for the components of each triangle's normal vector, and nine columns for the components of each triangle's vertices. For catalogs that contain a hierarchy of overlapping voids (V$^2_{\text{VIDE}}$), Tables \ref{tab:V2galviz} and \ref{tab:V2tri} contain entries for only the top-level ``parent'' voids.

Given that users of our void catalogs may wish to implement modified pruning procedures, we include the raw output of the ZOBOV algorithm as an additional void catalog. This V$^2_{\text{ZOBOV}}$ catalog contains the full hierarchy of voids defined in terms of zone linking and applies no minimum-size or central-density cuts. Tables \ref{tab:V2galviz} and \ref{tab:V2tri} are excluded from this catalog, as the single top-level ``parent'' void spans the entire survey volume.

The DESIVAST VACs are published online as part of DESI DR1 and can be accessed at \url{https://data.desi.lbl.gov/public/dr1/vac/dr1/desivast/}. The data points used for all figures in this paper are published online at \url{https://zenodo.org/records/22032815}. The pipeline used for this paper is available at \url{https://github.com/hbrincon/DESIVAST}.

Users of our catalogs can find tools for analyzing voids at \url{https://github.com/DESI-UR/VAST}. These tools include software for visualizing voids, measuring the volume fraction and galaxy membership within voids, and calculating void volumes. For users intending to create additional void catalogs with our void-finding software VAST, the software is also available at \url{https://github.com/DESI-UR/VAST}.

\bibliography{desi_bgs_voids_y1}{}
\bibliographystyle{aasjournal}

\end{document}